\documentclass[aps, prd, reprint, superscriptaddress]{revtex4-2}

\usepackage{amsmath,amssymb,calc, amsthm,bbm, epsfig,psfrag, mathtools,comment}

\usepackage{tikz}
\usepackage{tikz-cd}
\usepackage{tensor}

\usepackage{graphicx}
 \usepackage[all,cmtip]{xy}
\usepackage{dcolumn}
\usepackage{bm}

\newcommand{\overbar}[1]{\mkern 1.5mu\overline{\mkern-1.5mu#1\mkern-1.5mu}\mkern 1.5mu}

\begin{document}


\title{\boldmath Universality of Neural Network Field Theory}

\author{Christian Ferko}
\email{c.ferko@northeastern.edu}
\affiliation{The NSF Institute for Artificial Intelligence and Fundamental Interactions, and}
\affiliation{Department of Physics, Northeastern University, Boston, MA 02115, USA}

\author{James Halverson}
\email{j.halverson@northeastern.edu}
\affiliation{The NSF Institute for Artificial Intelligence and Fundamental Interactions, and}
\affiliation{Department of Physics, Northeastern University, Boston, MA 02115, USA}

\author{Aaron Mutchler}
\email{mutchler.a@northeastern.edu}
\affiliation{Department of Physics, Northeastern University, Boston, MA 02115, USA}

\date{\today}

\begin{abstract}
We prove that any quantum field theory, or more generally any probability distribution over tempered distributions in $\mathbb{R}^d$, admits a neural network description with a countable infinity of parameters. As an example, we realize the $2d$ Liouville theory as a neural network and numerically compute the three-point function of vertex operators, finding agreement with the DOZZ formula. 

\end{abstract}

\maketitle


\section{Introduction}

Quantum field theory (QFT) is a unifying framework which finds applications in several branches of modern physics, including particle physics, condensed matter systems, statistical physics, and quantum gravity. A recent proposal \cite{Halverson:2020trp,Halverson:2021aot} suggests that some QFTs can be \emph{defined} using neural networks (NNs), much as any continuous function can be approximated to arbitrary accuracy using a NN. This suggestion, referred to as the neural network - field theory (NN-FT) correspondence, has been extended to realize symmetries \cite{Maiti:2021fpy}; conformal field theories (CFTs) \cite{Halverson:2024axc} including conformal defects \cite{Capuozzo:2025ozt} and Virasoro symmetry \cite{Robinson:2025ybg}; fermions and supersymmetric theories \cite{Frank:2025zuk}; $\phi^4$ theory \cite{Demirtas:2023fir}; and worldsheet string theory \cite{Frank:2026bui}. For a summary of related ML results, see, e.g., \cite{Halverson:2024hax}.

It is natural to wonder about the generality of the NN-FT approach. For QFTs in one Euclidean spacetime dimension, i.e. ordinary theories of quantum mechanics (QM), under two mild assumptions it has been shown \cite{Ferko:2025ogz} that \emph{every} QM model admits a NN description with a countable infinity of parameters. These two assumptions are satisfied by any Euclidean QM model that obeys the Osterwalder-Schrader (OS) axioms, which guarantee that the theory can be analytically continued to yield a physical (e.g. unitary) theory in Lorentzian time.

However, QFTs in $1d$ are very special. To see why, suppose that we wish to define a QFT describing a scalar field $\phi : \mathbb{R}^d \to \mathbb{R}$ by taking the continuum limit of a lattice theory with spacing $a$. If the field has a kinetic term $\partial_\mu \phi \partial^\mu \phi$, its contributions to the action scale as
\begin{align}
    a^d \left( \frac{\phi ( x + a n ) - \phi ( x ) }{a} \right)^2 \, ,
\end{align}
where $n$ is a lattice vector. Assuming that $\phi$ is not constant, this term will remain finite as $a \to 0$ if $\phi ( x + a n ) - \phi ( x ) \sim a^{1 - d / 2}$. In $d = 1$, with a single coordinate $t$, this means that $\phi ( t + dt ) - \phi ( t ) \sim \sqrt{dt}$ for small $dt$, which is the characteristic scaling of Brownian motion. Thus in QM, the path integral measure is supported on Brownian-type paths which are continuous everywhere but differentiable nowhere. In higher dimensions, objects with this small-$a$ scaling are all subspaces of $\mathcal{S} ' ( \mathbb{R}^d )$, the space of tempered Schwartz distributions. Hence the field is not really a function which takes values, but it can be integrated against appropriate test functions, much like the Dirac delta distribution.

In constructive approaches to field theory, a QFT is therefore regarded as a probability distribution over Schwartz distributions obeying certain properties such as the OS axioms. In this work, we prove that every distribution over $\mathcal{S} ' ( \mathbb{R}^d )$ admits a NN description with a countable infinity of parameters, including cases corresponding to physical QFTs, thus generalizing the NN-QM result. In fact, at a formal level, every QFT can be realized as a NN with a \emph{single} parameter drawn from the uniform distribution on the unit interval $[0, 1]$. 

To frame our results in the most general setting, it will be convenient to define an abstract notion of \emph{generalized quantum system} (GQS), which is a random variable taking values in a space obeying a certain topological condition. Using measure theory, we prove that every GQS admits a countable NN representation, and finally we establish that every random variable taking values in $\mathcal{S}' ( \mathbb{R}^d )$ (e.g. a Euclidean QFT) is a GQS. The proof relies on the Borel isomorphism theorem, which guarantees the existence of an architecture for the construction.

In addition to demonstrating the existence of NN-FT, we wish to also show its utility. The specification of an architecture and parameter density may enable both analytic and numeric calculations. For instance, in implementations on a computer, one can often engineer a NN representation using orthonormal basis functions, retaining finitely many parameters. The resulting field configurations are ordinary functions rather than Schwartz distributions, but approach distributions as the number of included terms tends to infinity, just as one can approximate the delta distribution with a tall narrow Gaussian.

We illustrate the NN-FT approach by simulating Liouville theory on a sphere using a NN representation. This theory is interacting, has been proven to exist in a rigorous mathematical sense \cite{david2016liouville}, and admits an exact expression for the three-point function of vertex operators given by the DOZZ formula. We compare the NN-FT simulation of this three-point function to the exact results and find agreement to within a few percent error.

\section{Neural Network Quantum Mechanics}\label{sec:review}

Let us first review the $1d$ version of the NN-FT correspondence, which we call NN-QM, before generalizing. In quantum mechanics, we are instructed to sum over random paths $x(t)$ representing the position of a quantum particle. We refer to the domain of these random functions as the \emph{index set} $T$, and in QM the values $t \in T$ are interpreted as values of time, where $T = \mathbb{R}$ or $T = [ a, b ]$.

Consider the collection of random variables $\{ x ( t ) \mid t \in T \}$ which represent the position of a quantum particle at time $t$. For each finite collection of times $t_i \in T$, $i = 1 , \ldots, n$, suppose that we can assign a joint probability distribution function (PDF)
\begin{align}\label{output_joint_pdf}
    \mathbb{P} ( x ( t_1 ) , \ldots, x  ( t_n ) ) \, .
\end{align}
Such an assignment of a joint PDF to any finite collection of times is a special case of a \emph{stochastic process}. Abbreviating $x_i = x ( t_i )$, the density (\ref{output_joint_pdf}) defines correlators
\begin{align}\label{output_space_correlator}
    \langle x_1 \ldots x_n \rangle = \int d x_1 \ldots d x_n \, \mathbb{P} ( x_1 , \ldots, x_n ) \, x_1 \ldots x_n \, .
\end{align}
We refer to an expression of the form (\ref{output_space_correlator}) as the \emph{output-space} description of the correlation functions for such a stochastic process. The idea of NN-QM is to seek a dual \emph{parameter space} description of these correlation functions. In the $1d$ setting, such a description is defined by a parameterized family of random functions $\phi_\theta$ along with a joint PDF $P ( \theta )$. The functional form of $\phi_\theta ( t )$ is known as the architecture, which depends on a set of parameters $\theta$ and a time $t \in T$, and the PDF $P ( \theta )$ over these parameters is the parameter density. 
An example is a fully connected network with a single hidden layer,
\begin{align}\label{one_hidden_layer}
    \phi_\theta ( t ) = \sum_{i=1}^{N} w_i^{(1)} \sigma \left( w_i^{(0)} t + b_i^{(0)} \right) + b^{(1)} \, ,
\end{align}
where $\sigma$ is an activation function such as the sigmoid $\sigma ( y ) = \frac{1}{1 + e^{-y}}$, and the set of parameters $\theta \in \{ w_i^{(0)} , b_i^{(0)} , w_i^{(1)} , b^{(1)} \}$ consists of the weights and biases of the network, which are selected randomly according to some PDF. However, we will not restrict ourselves to functional forms like (\ref{one_hidden_layer}), instead referring to any parameterized family $\phi_\theta ( t )$ as a NN in this section.

Given such a NN architecture and parameter density, we can define correlation functions
\begin{align}\label{parameter_space}
    \langle \phi_\theta ( t_1 ) \ldots \phi_\theta ( t_n ) \rangle = \int d \theta \, P ( \theta ) \, \phi_\theta ( t_1 ) \ldots \phi_\theta ( t_n ) \, .
\end{align}
If the correlation functions (\ref{parameter_space}) coincide with (\ref{output_space_correlator}) for every finite collection of times $t_i$, we say $(\phi_\theta ( t ) , P ( \theta ) )$ is a \emph{neural network representation} for the stochastic process $x ( t )$, and we write $\phi_\theta ( t ) = x ( t )$. In this case, we say that the expression (\ref{parameter_space}) is a \emph{parameter-space} description of the correlation functions for the stochastic process.

Any tuple $(\phi_\theta ( t ) , P ( \theta ) )$ defines correlators (\ref{parameter_space}), thus (implicitly) joint PDFs (\ref{output_joint_pdf}), so every NN determines a stochastic process. Conversely, it is natural to ask whether every stochastic process determines a NN description (which need not be unique). As was pointed out in \cite{Ferko:2025ogz}, one can always choose a ``trivial'' NN representation: associate a separate parameter $\theta_t$ to each $t \in T$, and define $\phi_\theta ( t ) = \theta_t$ with the parameter density $P ( \theta )$ selected so that $P ( \theta_1 , \ldots , \theta_n ) = \mathbb{P} ( x_1 , \ldots, x_n )$ for any finite collection of $x_i$. This  representation requires uncountably many parameters if $T = \mathbb{R}$ or $T = [a, b]$.

A more interesting question is whether a stochastic process admits a NN representation with a countable or finite set of parameters $\theta_k$. For stochastic processes which correspond to physically reasonable quantum systems, a countable NN representation always exists. It was proven in \cite{Ferko:2025ogz} that, for any stochastic process on $T = [a, b]$ with finite and continuous two-point function $\langle x ( t ) x ( s ) \rangle$, there exists a ``canonical'' single-layer NN description
\begin{align}\label{kkl_decomposition}
    x ( t ) = \langle x ( t ) \rangle + \sum_{k=1}^{\infty} \theta_k f_k ( t ) \, ,
\end{align}
for certain functions $f_k ( t )$ and random variables $\theta_k$. Thus one can either perform a random draw of a path $x ( t )$, or equivalently draw countably many real parameters $\theta_k$ and form the sum (\ref{kkl_decomposition}), and these two procedures coincide.

\section{Universality Theorem}\label{sec:universality}

The fact that field configurations for QFTs in $d \geq 2$ are generically distributional is crucial for understanding several features that are absent in QM, such as coincident-point divergences in correlators. To accommodate this behavior, we will need to introduce a more general notion of stochastic process and apply some mathematical techniques from measure theory. This will allow us to prove that every probability distribution over $\mathcal{S}' ( \mathbb{R}^d )$, including Euclidean QFTs, admits a countable NN description.

Let $(\Omega, \mathcal{F}, \mathbb{P})$ be a probability space, where $\Omega$ is the sample space, $\mathcal{F}$ is the $\sigma$-algebra of measurable subsets of $\Omega$, and  $\mathbb{P}:\mathcal{F} \to [0,1]$ is a probability measure on $\mathcal{F}$. Let $(E, \mathcal{E})$ be a measurable space, where $E$ is the set of possible values and $\mathcal{E}$ is the $\sigma$-algebra of measurable subsets. We recall that a random variable $X$ is a measurable function $X : \Omega \to E$, which means that $X^{-1} ( A ) \in \mathcal{F}$ for every $A \in \mathcal{E}$. Let $T$ be an index set, which we will no longer assume is necessarily $\mathbb{R}$ or $[a, b]$ as above. The general definition of a stochastic process $x_t$ with state space $(E, \mathcal{E})$ is a family of random variables $\{ x_t \}_{t \in T}$, i.e. $x_t : \Omega \to E$ is a measurable function for each $t \in T$.

An equivalent definition involves the \emph{path space}
\begin{align}
    E^T = \{ f : T \to E \}
\end{align}
of all functions from the index set $T$ into $E$. As $(E, \mathcal{E})$ is a measurable space, we may endow $E^T$ with the product $\sigma$-algebra $\mathcal{E}^{\otimes T}$ in order to construct the measurable space $(E^T, \mathcal{E}^{\otimes T})$. Then a stochastic process is a random variable $\mathbf{x} : \Omega \to E^T$, i.e. a measurable function taking values in the path space $E^T$. Here we use the boldface symbol $\mathbf{x}$ to emphasize that draws of this random variable $\mathbf{x}$ represent entire sample paths, rather than values $x_t$ of the random variable at particular points $t \in T$. In the scalar field case they might conventionally be called $\phi$; we instead use $\textbf{x}$ to generalize the QM case $x_t$, and emphasize that $\textbf{x}$ is not a spacetime coordinate.

This path space formulation can be related to the preceding definition as follows. Let $\pi_t : E^T \to E$ be the evaluation map which acts as $\pi_t ( f ) = f ( t )$ on any function $f : T \to E$. By the properties of the product $\sigma$-algebra, since $\mathbf{x}$ is a measurable function, each individual random variable $x_t = \pi_t \circ \mathbf{x}$ is a measurable function from $\Omega$ to $E$. In particular, for any finite collection $t_1 , \ldots, t_n \in T$, the vector of random variables $\left( x_{t_1} , \ldots, x_{t_n} \right) : \Omega \to E^n$ is also measurable, which means that we can assign a joint probability measure (\ref{output_joint_pdf}) to this collection of outputs.

As before, we now discuss the dual parameter-space description. A stochastic process $\mathbf{x}$ from a probability space $(\Omega, \mathcal{F}, \mathbb{P})$ into the path space $(E^T, \mathcal{E}^{\otimes T})$ is said to admit a neural network representation, with parameters valued in a measurable space $( \Upsilon, \mathcal{G})$, if there exist measurable maps $\Theta : \Omega \to \Upsilon$ and $\phi_\theta : \Upsilon \to E^T$ such that
\begin{align}\label{x_factorize}
    \mathbf{x} = \phi_\theta \circ \Theta \, .
\end{align}
We refer to $\Theta$ as the parameter map and $\phi_\theta$ as the architecture. The pushforward of the measure $\mathbb{P}$ on $(\Omega, \mathcal{F}, \mathbb{P})$, along the measurable map $\Theta$, defines a measure $P = \Theta_\ast \mathbb{P}$ on $(\Upsilon, \mathcal{G})$ which is called the parameter measure. Let $\mathcal{B} ( \mathbb{R}^n )$ be the Borel $\sigma$-algebra on $\mathbb{R}^n$, and likewise let $\mathcal{B} ( \mathbb{R} )^{\otimes \mathbb{N}}$ be the corresponding Borel $\sigma$-algebra on sequences in $\mathbb{R}^{\mathbb{N}}$. If
\begin{align}
    ( \Upsilon, \mathcal{G} ) = \left( \mathbb{R}^n, \mathcal{B} ( \mathbb{R}^n ) \right) \text{ or  } ( \Upsilon, \mathcal{G} ) = \left( \mathbb{R}^{\mathbb{N}}, \mathcal{B} ( \mathbb{R} )^{\otimes \mathbb{N}} \right) \, , 
\end{align}
then we say $\mathbf{x}$ admits a NN representation with finitely many or countably many parameters, respectively.

This condition can be stated in more categorical terms: in the category \textbf{Meas} of measurable spaces, where morphisms are measurable maps, the diagram 
\[\begin{tikzcd}
	{(\Omega, \mathcal{F})} && {( E^T, \mathcal{E}^{\otimes T} )} & {} \\
	\\
	\\
	{( \Upsilon , \mathcal{G} )}
	\arrow["\mathbf{x}", from=1-1, to=1-3]
	\arrow["\Theta"', from=1-1, to=4-1]
	\arrow["{\phi_\theta}"', dashed, from=4-1, to=1-3]
\end{tikzcd}\]
commutes. The existence of a NN representation for a stochastic process $\mathbf{x}$ is the statement that this random variable factors through another measurable space $( \Upsilon, \mathcal{G} )$ of parameters $\theta$, and again we typically take either $\Upsilon = \mathbb{R}^n$ with finitely many parameters $\{ \theta_1 , \ldots, \theta_n \}$ or $\Upsilon = \mathbb{R}^{\mathbb{N}}$ with countably many parameters $\{ \theta_i \mid i \in \mathbb{N} \}$. For fixed parameters $\theta$, the input to the NN is an element $t \in T$ and the output of the NN is the evaluation $\pi_t \circ \phi_\theta$.

\medskip
We may now state some of the observations of Section \ref{sec:review} in this more general setting. In the special case of stochastic processes $x_t : T \to \mathbb{R}$ with $T = \mathbb{R}$ or $T = [ a, b ]$, we commented that one can always choose a ``trivial'' NN representation. The same is true here: we simply choose $( \Upsilon , \mathcal{G} ) = ( E^T, \mathcal{E}^{\otimes T} )$ with the maps $\Theta = \mathbf{x}$ and $\phi_\theta = \mathbb{I}$. Then there is one parameter $\theta_t = \pi_t \circ \mathbf{x} = x_t$ for each value $t$ in the index set, which in most cases of interest (such as QFT in the continuum) is uncountably infinite. Furthermore, as in NN-QM, beginning from a fixed probability space $(\Omega, \mathcal{F}, \mathbb{P})$, every tuple $( \Theta, \phi_\theta )$ determines a stochastic process $\mathbf{x}$ by the composition of maps (\ref{x_factorize}).

Again, one might ask when a general stochastic process admits a NN representation with countably many parameters, as in (\ref{kkl_decomposition}). To guarantee this, we must impose some additional topological structure. Suppose that the process $\mathbf{x}$ takes values almost surely in an uncountable subspace $\mathcal{X} \subseteq E^T$ such that $( \mathcal{X} , \mathcal{B} ( \mathcal{X} ) )$ is a \emph{standard Borel space}. We then regard $\mathbf{x}$ as a random variable from $\Omega$ to $\mathcal{X}$, which means that it is measurable with respect to the Borel $\sigma$-algebra $\mathcal{B} ( \mathcal{X} )$. To ensure this condition, it is sufficient (but not necessary) to assume that $\mathcal{X}$ is a \emph{Polish space}, i.e. a separable completely metrizable topological space. This assumption is satisfied by spaces of interest for QM like the space of square-integrable functions $L^2 ( [a, b] )$, but not by the space $\mathcal{S}' ( \mathbb{R}^d )$ relevant for QFT (which is not even metrizable). Despite this, $\mathcal{S}' ( \mathbb{R}^d )$ is still standard Borel --- in fact it is Borel-isomorphic to a Borel subset of a Polish space --- as we will see.

Assuming that $\mathbf{x}$ takes values in a standard Borel space $( \mathcal{X} , \mathcal{B} ( \mathcal{X} ) )$ with uncountable $\mathcal{X}$, we may invoke the Borel isomorphism theorem, which guarantees that any two uncountable standard Borel spaces are Borel isomorphic. In particular, $\mathcal{X}$ is isomorphic to the space $\mathbb{R}^{\mathbb{N}}$, i.e. the space of infinite sequences of parameters $\{ \theta_1 , \theta_2 , \ldots \}$. Note that although each element of $\mathbb{R}^{\mathbb{N}}$ is a sequence of countably many parameters, the space $\mathbb{R}^{\mathbb{N}}$ of \emph{all} such sequences is uncountable (even the space of all allowed initial entries $\theta_1$ in the sequence is $\mathbb{R}$, which is uncountable). Explicitly, there exists a Borel isomorphism 
\begin{align}
    \psi : \mathcal{X} \to \mathbb{R}^{\mathbb{N}} \, ,
\end{align}
which means that $\psi$ is bijective, and both $\psi$ and $\psi^{-1}$ map Borel sets to Borel sets. We may then construct a neural network representation with countably many parameters by choosing the parameter space $( \Upsilon , \mathcal{G} ) = ( \mathbb{R}^{\mathbb{N}} , \mathcal{B} ( \mathbb{R} )^{\otimes \mathbb{N}} )$. The parameter map 
is $\Theta = \psi \circ \mathbf{x}$, and the architecture is $\phi_\theta = \psi^{-1}$, yielding the factorization
\begin{align}
    \Omega \overset{\Theta}{\longrightarrow} \mathbb{R}^{\mathbb{N}} \overset{\phi_\theta}{\longrightarrow} \mathcal{X} \subseteq E^T \, .
\end{align}
The decomposition (\ref{kkl_decomposition}) in NN-QM corresponds to a ``preferred'' isomorphism based on the covariance structure of the underlying process, whose special properties are not implied by this argument. However, this general measure-theoretic result ensures that any stochastic process valued in a standard Borel space admits a NN representation with countably many parameters, even if there is no distinguished ``canonical'' representation like (\ref{kkl_decomposition}).

We use the term \emph{generalized quantum system} for any stochastic process $\mathbf{x}$ which takes values almost surely in a standard Borel space $( \mathcal{X} , \mathcal{B} ( \mathcal{X} ) )$ where $\mathcal{X}$ is uncountable, regardless of whether $\mathcal{X} \subseteq E^T$ for some $E$ and $T$. We have therefore extended the result of \cite{Ferko:2025ogz} to the statement that any GQS admits a neural network representation with countably many parameters.

\medskip
\textbf{Universality Result}. Our main result is that this general argument applies to the setting relevant for conventional QFTs. 
Let $\mathcal{S} ( \mathbb{R}^d )$ be the Schwartz space of test functions on $\mathbb{R}^d$, and let $\mathcal{S} ' ( \mathbb{R}^d )$ be its dual  which is the space of tempered distributions. It is well-known (see, for instance, \cite{hida2013white}) that $\mathcal{S} ' ( \mathbb{R}^d )$ is a Lusin space, which means that it is the image of a Polish space under a continuous bijection. Furthermore, for any Lusin space $L$, it is known that $( L , \mathcal{B} ( L ) )$ is standard Borel (see e.g. the appendix of \cite{dedecker2006parametrized}). Therefore, a stochastic process $\mathbf{x}$ taking values in the state space $\left( \mathcal{S} ' ( \mathbb{R}^d ) , \mathcal{B} \left( \mathcal{S} ' ( \mathbb{R}^d ) \right) \right)$ satisfies the definition of a GQS, and thus admits a NN description with a countable infinity of parameters. Thus if we regard a QFT as a random variable which takes values in the space of tempered distributions on $\mathbb{R}^d$, perhaps obeying additional axioms like OS, every QFT admits a countable or even finite NN description.

The above proof is quite general, but non-constructive. In specific cases of interest, one can use an explicit isomorphism between $\mathcal{S}'$ and the space of sequences $( \theta_1, \theta_2 , \ldots )$ of at most polynomial growth (see, e.g., Chapter 8 of \cite{kupiainen2014rg}). For instance, choosing a basis of Hermite functions $\{ \varphi_0, \varphi_1 , \ldots \}$ on $\mathbb{R}$, every tempered distribution $T \in \mathcal{S}' ( \mathbb{R} )$ acts on functions $f \in \mathcal{S} ( \mathbb{R} )$ as
\begin{align}
    T ( f ) = \sum_{k=1}^{\infty} y_k \langle f, \varphi_{k-1} \rangle \, , \qquad y_k = T ( \varphi_{k-1} ) \, ,
\end{align}
and is thus determined by a sequence $y_k$ that can be identified with suitable random parameters $\theta_k$. This construction extends straightforwardly to $\mathcal{S}' ( \mathbb{R}^d )$. A similar basis will be used on the $2$-sphere in Section \ref{sec:liouville}.

Another type of GQS involves variables $\mathbf{x}$ which take values in continuous functions, as in NN-QM, but with index set $T = \mathbb{R}^d$. 
Such a stochastic process can be viewed as a theory of random continuous functions $\phi ( x_1 , \ldots, x_d )$ in $d$ Euclidean dimensions. However, as we reviewed, such random functions do not appear to be the correct stochastic processes for describing QFTs in $d \geq 2$ dimensions; the path integral measure for QFTs is supported on proper distributions rather than continuous functions, at least for scalar theories with a kinetic term obtained from continuum limits of lattice theories. One should invoke the preceding result for $\mathcal{S} ' ( \mathbb{R}^d )$ in such cases.


It is natural to ask whether a GQS admits a NN representation with a \emph{finite} number of parameters. Formally, the answer is yes, and only one is required. Every standard Borel space is either countable or Borel-isomorphic to the interval $[0,1]$ with the standard Borel $\sigma$-algebra, and as we have assumed $\mathcal{X}$ is uncountable, every GQS admits a neural network description with one parameter $\theta \in [0 , 1]$. An intuitive way to see this is as follows. Given an infinite-precision draw of $\theta$ from, say, the uniform distribution $U ( [0, 1] )$, construct its decimal expansion, and partition the decimals into a countably infinite set of independent expansions by a Cantor-diagonalization-like procedure. This gives countably many independent draws from $U ( [ 0 , 1 ] )$. Draws from an arbitrary joint PDF $P ( \theta_1, \theta_2 , \ldots )$ on countably many random variables can then be engineered from applying deterministic functions to these countably many random uniform variables, using Copula-like methods or an inverse Rosenblatt transformation. Thus, formally, one draw from a random uniform is enough to sample any realization of a GQS. 

However, for restricted architectures -- such as finite linear combination of functions with random coefficients -- one can sometimes prove no-go theorems forbidding finite NN representations of physical quantum systems, as will be discussed in \cite{toappear} for NN-QM using the K\"all\'en-Lehmann spectral representation. This result can be phrased as the statement that a finite fixed-features model, with features $f_k ( t )$, cannot reproduce a unitary QM model. Deep architectures correspond to choosing a more complicated map $\phi_\theta : \mathbb{R} \to \mathbb{R}$ to represent a quantum system, such as modifying (\ref{one_hidden_layer}) to include additional layers with successive applications of the activation $\sigma$.

\section{Liouville Theory}\label{sec:liouville}

The Liouville CFT finds many applications in physics, such as $2d$ quantum gravity and noncritical strings \cite{POLYAKOV1981207,Seiberg:1990eb}, and in the study of $2d$ critical phenomena on random surfaces \cite{doi:10.1142/S0217732388000982,DISTLER1989509}, while related ideas have been applied in studies of turbulence \cite{Mandelbrot_1974,rhodes2014gaussian}. One can view this model as an effective theory for the conformal mode $\phi$ of a $2d$ metric $g_{\mu \nu} = e^{2 \phi} \hat{g}_{\mu \nu}$ when performing the path integral over metrics. As our discussion has been fairly rigorous, we find it instructive to apply the NN-FT approach to Liouville theory, since advances in the probability literature \cite{Hoegh-Krohn:1971xiq,kahane1985chaos,david2016liouville,kupiainen2020integrability} have shown that Liouville theory is well-defined at a formal mathematical level using the theory of Gaussian multiplicative chaos (GMC). See \cite{Chatterjee:2024phq} for a review of such approaches. The probabilistic definition of Liouville theory parallels its NN-FT realization, with a deformed parameter density that implements GMC.

Classically, Liouville theory is described by the action
\begin{align}\label{liouville_action}
    S = \int_{\Sigma} d^2 x \, \sqrt{g} \, \left( \frac{1}{4 \pi} \partial^\mu \phi \partial_\mu \phi + \frac{Q}{4 \pi} R \phi + \mu e^{2 b \phi} \right) \, ,
\end{align}
where $b$ and $\mu$ are parameters, $Q = b + \frac{1}{b}$, and $R$ is the Ricci scalar of the surface $\Sigma$ on which the theory is defined. We take $\Sigma = S^2$, where $R$ is constant. 

We are interested in correlation functions involving the vertex operator $V_\alpha = : e^{2 \alpha \phi ( x )} :$. The three-point function of such operators takes the form
\begin{align}\label{three_point_vertex}
    \langle V_{\alpha_1} ( x_1 ) V_{\alpha_2} ( x_2 ) V_{\alpha_3} ( x_3 ) \rangle = \frac{C ( \alpha_1, \alpha_2, \alpha_3 )}{| x_{12} |^{2 \Delta_{123}} | x_{13} |^{2 \Delta_{132}} | x_{23} |^{2 \Delta_{231}}} \, ,
\end{align}
where $x_{ij} = x_i - x_j$, $\Delta_{ijk} = \Delta_i + \Delta_j - \Delta_k$, and $\Delta_i = \alpha_i ( Q - \alpha_i )$. The quantity $C ( \alpha_1, \alpha_2, \alpha_3 )$ is provided in terms of certain special functions by the DOZZ formula \cite{Dorn:1994xn,Zamolodchikov:1995aa}; its explicit form is somewhat unwieldy, but can be found in Appendix \ref{sec:dozz}.

Our goal is to realize Liouville as an NN-FT and reproduce the function $C$ in (\ref{three_point_vertex}). To do this, we must specify an architecture $\phi_\theta$ and a parameter density $P ( \theta )$. We begin by splitting the field $\phi_\theta$ into a zero mode contribution $c$ and a term $X_a$ involving real spherical harmonics,
\begin{align}\label{split_zero}
    \phi_\theta = c + X_{a} = c + \sum_{\ell=1}^{L} \sum_{m = - \ell}^{\ell} a_{\ell, m} Y_{\ell, m} \, .
\end{align}
The parameters $\theta$ of the network are the value $c$ of the zero mode, along with the set of coefficients $a_{\ell, m}$. Following the strategy described in \cite{Demirtas:2023fir}, we first endow $X_{a}$ with the architecture of a free field on $S^2$, which corresponds to drawing the $a_{\ell, m}$ from independent Gaussians
\begin{align}\label{alm_distribution}
    a_{\ell, m} \sim \mathcal{N} \left( 0, \sigma^2 = \frac{4 \pi}{\ell ( \ell + 1 ) } \right) \, ,
\end{align}
and later incorporate interactions through a deformation of the parameter density. As usual, for $L \to \infty$, the sum defining $X_{a}$ does not converge to a well-behaved function, but it converges in the sense of distributions. This is a different notion of an $L \to \infty$ limit than the infinite-width limit which defines NN Gaussian processes \cite{neal}; although the $a_{\ell, m}$ are Gaussian variables, we emphasize that the full field $\phi_\theta$ is interacting even for $L \to \infty$.

In the path-integral description of Liouville correlators, the zero mode $c$ enters in three distinct places: (i) the curvature coupling $\int_{S^2} \sqrt{g} \frac{Q}{4 \pi} R c = 2 Q c$, (ii) the vertex operators $V_\alpha ( x ) = e^{2 \alpha c} : e^{2 \alpha X ( x )} : $, and (iii) in the exponential interaction term, which involves both $c$ and $X$. We engineer these contributions using an effective conditional parameter density for the zero mode, or more precisely for the variable $t = e^{2 b c}$. The contribution (iii) to the exponential from the non-zero-mode $X_{a}$ will introduce the GMC factor into this parameter density. The full parameter density takes the conditional form
\begin{align}
    P ( \theta ) = P ( t, a_{\ell, m} ) = P ( t \mid a_{\ell, m} ) \cdot P ( a_{\ell, m} ) \, ,
\end{align}
where $a_{\ell, m}$ schematically denotes the set of all $a_{\ell, m}$ for $\ell \geq 1$, and where the zero mode density requires conditioning on the non-zero modes in a way which describes the interactions. In fact, the integral over $t$ can be evaluated exactly in closed form, but for completeness we also simulate the zero mode contribution, to obtain a fully ``first-principles'' Monte Carlo NN-FT simulation. We do this using the un-normalized conditional distribution
\begin{align}\label{P_un_normalized}
    P ( t \mid a_{\ell, m} ) = \frac{1}{2b} t^{s-1} \exp \left( - M_X ( a_{\ell, m} ) t \right) \, ,
\end{align}
where $s = \frac{\sum \alpha_i - Q}{b}$, which involves a numerically estimated and normal-ordered version of the GMC factor
\begin{align}
    M_X = \mu \int_{S^2} d^2 x \, \sqrt{g} \, : e^{2 b X_a} \, : \, ,
\end{align}
by first drawing the $a_{\ell, m}$, computing $M_X$ which implicitly depends on the $a_{\ell, m}$ through $X_a$, and performing conditional draws of $t$, and finally accounting for the normalization factor. Further details about our implementation can be found in Appendix \ref{sec:implementation}.

\begin{figure}
  \centering
  \includegraphics[width=\linewidth]{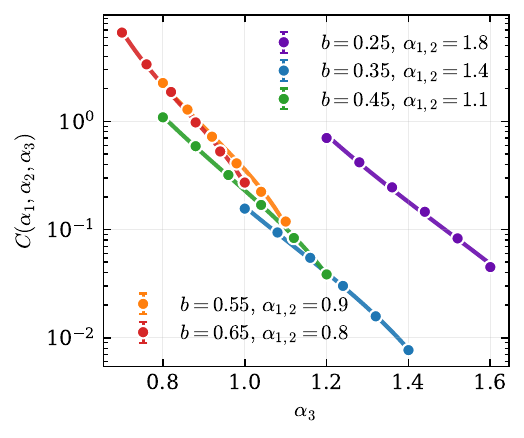}
  \caption{The three-point function (\ref{three_point_vertex}) is computed for various values of $b$, with $\alpha_1 = \alpha_2$ fixed, and as $\alpha_3$ is varied. The NN-FT simulation is performed with $L = 30$ spherical harmonics and $10$ experiments with $50,000$ runs per experiment; error bars are shown measuring the variance across experiments but are too small to be seen. A UV regulator is introduced by pixelating the sphere at $100$ points of latitude and $200$ points of longitude. Exact results from the DOZZ formula are shown in solid curves and lie within all error bars.}
  \label{fig:dozz_prl_plot}
\end{figure}

After drawing random $a_{\ell, m}$ and $t = e^{2 b c}$, we evaluate an average of appropriate normal-ordered three-point functions of vertex operators (\ref{three_point_vertex}), 
sampled at a fixed choice of the three points $x_i$ (taken to be the north pole, south pole, and one point on the hemisphere of $S^2$). We then fix one undetermined overall $\alpha$-independent prefactor using a numerical fit and compare the results to the DOZZ formula. The results appear in Figure \ref{fig:dozz_prl_plot}.

\section{Conclusion}\label{sec:conclusion}

In this work, we have proven that every Euclidean QFT admits a NN description with a countable infinity of parameters -- or, indeed, with a \emph{single} parameter -- generalizing the universality result of \cite{Ferko:2025ogz} for NN-QM. As an example, we realized the $2d$ Liouville theory as a NN-FT and reproduced the prediction of DOZZ for the three-point function of vertex operators on a sphere.

Notably, since the Borel isomorphism theorem ensures the existence of a NN description of a QFT, it has an architecture that is bijective between parameters and Schwartz distributions. This invertibility of the architecture between parameters and $\mathcal{S}'(\mathbb{R}^d)$ is not a general feature of NN-FT, and indeed many NN-FT descriptions enable concrete calculations of physical quantities, but with a non-invertible architecture. For instance, the free scalar architecture of \cite{Halverson:2021aot} is non-invertible but may be restricted to the free boson in 2d and utilized to construct the bosonic string \cite{Frank:2026bui} and compute string amplitudes.

There remain several interesting directions for future research. Although we established that every GQS admits a NN description, including the distributions $\mathcal{S}' ( \mathbb{R}^d )$ that provide the foundation of constructive field theory, it is not clear which NN-FTs obey appropriate physical conditions, such as the OS axioms. In $1d$, some mechanisms are known \cite{Ferko:2025ogz} for engineering reflection positivity, perhaps the most subtle of the OS axioms. It would be exciting to establish similar results for NN-FTs in $d \geq 2$.

Another avenue for investigation is to engineer other target theories as NN-FTs. For instance, $\phi^4$ theory in $3d$ is also known to exist at a rigorous mathematical level by virtue of its super-renormalizability \cite{Glimm:1987ng}, and it would be interesting to realize this model numerically via a NN, perhaps using the $\phi^4$ treatment of \cite{Demirtas:2023fir}. It would also be intriguing to realize integrable models, such as sine-Gordon theory, since some of their observables admit exact  expressions that can be compared to numerics.

Finally, we have only discussed ensembles of NNs with random parameters and have not mentioned training. Rather than guessing an architecture and parameter density to engineer a target QFT, it would be very useful if one could instead train a NN to reproduce a given theory of interest. Some preliminary comments describing how one might approach this for $1d$ models appeared in \cite{Ferko:2025ogz}. In our context, the invertibility of the architecture associated to the Borel isomorphism and the existence of non-invertible architectures demonstrates the existence of \emph{overparameterized} NN-FTs, ones where the architecture is many-to-one and has more parameters than strictly required. It would be interesting to explore whether this phenomenon gives rise to improved training of NN-FTs akin to overparametrization and double descent \cite{doi:10.1073/pnas.1903070116} in ML. It may be advantageous to implement this approach for higher-dimensional field theories, facilitating the use of training to explore the space of QFTs.

\begin{acknowledgments}
We are grateful to Ning Bao, Vishnu Jejjala, Sarah Harrison, and Fabian Ruehle for helpful comments. This work is supported by the National Science Foundation under Cooperative Agreement PHY-2019786 (the NSF AI Institute for Artificial Intelligence and Fundamental Interactions). J.H. is supported by NSF grant PHY-2209903. A.\,M. is supported by the National Science Foundation grant PHY-2209903.
\end{acknowledgments}

\bibliographystyle{apsrev4-1} 
\bibliography{apssamp}

\onecolumngrid

\appendix
\setcounter{section}{0}
\setcounter{equation}{0}
\newpage

\begin{center}{\large \textbf{SUPPLEMENTAL MATERIAL
\\\vspace{0.1cm}
 (APPENDICES)
}
}
\end{center}

In the Supplemental Material accompanying our letter, we provide additional information about the $2d$ Liouville theory in Appendix \ref{sec:dozz} and further details about our numerical implementation of Liouville using NN-FT in Appendix \ref{sec:implementation}. Neither of these Appendices are essential to the content in the main body of this letter. We have primarily chosen to relegate certain cumbersome formulas to Appendix \ref{sec:dozz} in order to avoid cluttering the body of this article, and to describe implementation details in Appendix \ref{sec:implementation} for the benefit of readers who may wish to reproduce our results.

\section{Liouville and the DOZZ Formula}\label{sec:dozz}

One of the many reasons that the Liouville conformal field theory (LCFT) is of interest is because, when the Polyakov path integral over metrics was introduced in \cite{POLYAKOV1981207}, it was found that the effective theory of the conformal mode for the metric is described by LCFT. This conformal mode can be viewed as a scalar field $\phi$ and the dynamics of this scalar are described by the action (\ref{liouville_action}) given in the body of this letter. The basic local primary operators of interest in LCFT are the vertex operators
\begin{align}\label{vertex_operator}
    V_\alpha ( x ) = \; : e^{2 \alpha \phi ( x ) } : \, ,
\end{align}
which have conformal weights
\begin{align}
    \Delta_\alpha = \overbar{\Delta}_\alpha = \alpha ( Q - \alpha ) \, ,
\end{align}
with $Q = b + \frac{1}{b}$ as in the main text. 
The central charge of this theory is $c_L = 1 + 6 Q^2$. 

It is believed that the entire quantum theory of LCFT is completely determined by the three-point function of vertex operators on the sphere, due to bootstrap arguments. 
Formally, this three-point function is given by the path integral expression
\begin{align}\label{three_point_function}
    \langle V_{\alpha_1} ( x_1 ) V_{\alpha_2} ( x_2 ) V_{\alpha_3} ( x_3 ) \rangle = \int \mathcal{D} \phi \, e^{- S} V_{\alpha_1} ( x_1 ) V_{\alpha_2} ( x_2 ) V_{\alpha_3} ( x_3 ) \, .
\end{align}
We say ``formally'' because, as usual in interacting QFTs, one must first make sense of the weight $\mathcal{D} \phi \, e^{-S}$ in (\ref{three_point_function}) in order for this quantity to have a rigorous mathematical meaning. For LCFT, this can be achieved \cite{david2016liouville}, although historically it was not how the correct expression for the three-point function (\ref{three_point_function}) was first discovered. The coordinate dependence of this correlation is fixed to the form (\ref{three_point_vertex}) by conformal invariance, and the remaining undetermined factor $C ( \alpha_1 , \alpha_2 , \alpha_3 )$ was proposed independently by Dorn-Otto \cite{Dorn:1994xn} and $\left( \text{Zamolodchikov} \right)^2$ \cite{Zamolodchikov:1995aa} to take the form
\begin{align}\label{DOZZ_C}
    C ( \alpha_1, \alpha_2, \alpha_3 ) &= \left( \pi \mu \gamma ( b^2 ) b^{2 - 2 b^2} \right)^{-s} \nonumber \\
    &\qquad \cdot \frac{\Upsilon_b ' ( 0 ) \Upsilon_b ( 2 \alpha_1 ) \Upsilon_b ( 2 \alpha_2 ) \Upsilon_b ( 2 \alpha_3 ) }{\Upsilon_b ( \alpha_1 + \alpha_2 + \alpha_3 - Q ) \Upsilon_b ( \alpha_1 + \alpha_2 - \alpha_3 ) \Upsilon_b ( \alpha_2 + \alpha_3 - \alpha_1 ) \Upsilon_b ( \alpha_1 + \alpha_3 - \alpha_2 ) } \, ,
\end{align}
for all $\alpha_1, \alpha_2, \alpha_3 \in \mathbb{C}$ (however, in what follows, we always take the $\alpha_i$ to be real, as is done in the probability literature). Here $\gamma ( x )$ is given in terms of a ratio of standard gamma functions by
\begin{align}
    \gamma ( x ) = \frac{\Gamma ( x )}{\Gamma ( 1 - x )} \, ,
\end{align}
and $\Upsilon_b ( z )$ can be defined for $0 < \mathrm{Re} ( z ) < Q$ by the integral formula
\begin{align}
    \log \left( \Upsilon_b ( z ) \right) = \int_0^\infty \frac{dt}{t} \left( \left( \frac{Q}{2} - z \right)^2 e^{-t} - \frac{ \left( \sinh \left( \left( \frac{Q}{2} - z \right) \frac{t}{2} \right) \right)^2}{\sinh \left( \frac{b t}{2} \right) \sinh \left( \frac{t}{2b} \right) } \right) \, . 
\end{align}
Finally, $s = \frac{\sum_i \alpha_i - Q}{b}$, as in the body. Although the formula (\ref{DOZZ_C}) was not rigorously derived by DOZZ, instead being proposed based on an analytic continuation argument, it has since been proven using probability techniques \cite{kupiainen2020integrability}.

We have emphasized that typical field configurations $\phi$ for QFTs in $d \geq 2$ are distributions rather than functions, and hence their products and exponentials are ill-defined. Therefore, quantities such as the Liouville interaction $e^{2 b \phi}$ must be renormalized. What is very special about Liouville theory is that renormalization of exponentials of the field $\phi$ is rather simple: it can be implemented simply by normal ordering with respect to a \emph{free} field measure, and this renormalization introduces a positive multiplicative factor (which is related to the so-called Gaussian multiplicative chaos of Kahane \cite{kahane1985chaos}, as we have mentioned).

One way to understand this is the following argument, nicely reviewed in \cite{Chatterjee:2024phq}, to which we refer the reader for further details. We write $: Z :$ for the normal-ordering of a random variable $Z$. For a mean-zero Gaussian random variable $Z$, one has 
\begin{align}\label{normal_order_gaussian}
    : \exp ( Z ) : = \exp \left( Z - \frac{1}{2} \mathbb{E} [ Z^2 ] \right) \, .
\end{align}
The field $X$ of LCFT, after separating the zero mode by writing
\begin{align}
    \phi = c + X \, ,
\end{align}
as in (\ref{split_zero}), is not such a Gaussian random variable, but rather a random Gaussian distribution. This means that for any appropriate test function $f$, the integral
\begin{align}
    X ( f ) = \int_{\Sigma} d^2 x \, \sqrt{g} \, X ( x ) f ( x ) 
\end{align}
is a Gaussian random variable. Therefore, although strictly speaking we cannot define the normal-ordered quantity
\begin{align}\label{normal_order_gaussian_X}
    : \exp ( X ) : \, \neq \, \exp \left( X - \frac{1}{2} \mathbb{E} [ X^2 ] \right)
\end{align}
as $\mathbb{E} [ X^2 ]$ is the coincident-point limit of the field's two-point function and therefore infinite, we can construct the smeared field
\begin{align}
    X_\epsilon ( x ) = X ( f_{\epsilon,x} ) = \int_{\Sigma} d^2 y \, \sqrt{g} X ( y ) f_{\epsilon, x} ( y ) 
\end{align}
where $f_{\epsilon, x} ( y )$ is referred to as a mollifier. Several choices of mollifier are possible, but for concreteness, one can think of $f_{\epsilon, x}$ as a Gaussian function of characteristic width $\epsilon$ centered at the point $x$. 

Because $X_\epsilon ( x )$ is now a mean-zero Gaussian random variable, one can define the normal-ordered quantity
\begin{align}\label{normal_ordered_X}
    : \exp \left( k X_{\epsilon} ( x ) \right) : = \exp \left( k X_{\epsilon} ( x ) - \frac{k^2}{2} \mathbb{E} \left[ X_{\epsilon} ( x )^2 \right] \right) \, ,
\end{align}
and ultimately take the limit $\epsilon \to 0$. This is the procedure used to define renormalized vertex operators $V_\alpha$ when $k = 2 \alpha$, or the renormalized interaction term in LCFT when $k = 2 b$.

Our discussion of Liouville theory here is far from complete, as we primarily wish to motivate the quantity (\ref{DOZZ_C}) which we will compute numerically using NN-FT. For a more comprehensive introduction, see any of the articles \cite{Seiberg:1990eb,Teschner:2001rv,Nakayama:2004vk,vargas2017lecture,Chatterjee:2024phq,rhodes2025two} and references therein.

\section{Details of NN-FT Implementation}\label{sec:implementation}

Let us now explain how LCFT is realized as a NN-FT and how the three-point function of vertex operators on the sphere is computed numerically.

As an initial remark, we note that there are actually two distinct ways in which one can perform this calculation. The first way relies upon a result known as the Girsanov theorem in probability theory, and can be viewed as a form of completing the square in the expectation value (\ref{three_point_function}). This technique shifts the value of the field $X$ as
\begin{align}
    X ( x ) \longrightarrow X ( x ) + 2 \sum_{i=1}^{3} \alpha_i G^{(2)} ( x, x_i ) \, ,
\end{align}
where $G^{(2)} ( x, y )$ is the free propagator. Such a shift has the effect of eliminating the vertex operators in the three-point function, and reduces the numerical computation of the three-point function to an estimate of a certain power of the normal-ordered Liouville interaction term, dressed by appropriate factors. One can carry out this estimate using a NN-FT procedure that follows most of the steps that we will describe shortly, except for the treatment of the vertex operators. In fact, we have implemented this procedure and verified that it yields a match to the DOZZ formula of a comparable quality to the one reported in the body of this letter while using less compute.

The reason that we have not emphasized this method, preferring instead to present the more explicit approach below, is the following. Although the Girsanov shift is quite elegant, and connects to ideas of Goulian-Li \cite{PhysRevLett.66.2051} and the Coulomb gas formalism, it relies upon path integral techniques in intermediate steps in order to simplify the form of the correlation functions. Of course, these techniques are correct and can be made mathematically rigorous, but the resulting quantities which one computes after the shift are not \emph{literal} correlation functions of the form (\ref{parameter_space}) as defined in the NN-FT paradigm. The use of these results might be seen as a crutch, whereas one could argue that a fully ``first-principles'' NN-FT implementation should genuinely sample values of appropriate parameters and then compute an ensemble average of operators defining the correlation functions of interest.

We now describe how this latter, more direct, procedure is carried out. As we mentioned around equation (\ref{split_zero}) in the body, we first choose an architecture $\phi_\theta = c + X_a$ which separates the zero-mode contribution $c$, and then expand the non-zero-mode term $X_a$ in real spherical harmonics. Choosing independent Gaussian coefficients $a_{\ell, m}$ as in (\ref{alm_distribution}) generates approximate draws of a free scalar field on $S^2$. For any finite $L$ and test function $f$, the pairing
\begin{align}\label{Xa_gaussian_sum}
    X_a ( f ) = \sum_{\ell = 1}^{L} \sum_{m = - \ell}^{\ell} \int_{S^2} d^2 x \, \sqrt{g} \, a_{\ell, m} Y_{\ell, m} ( x ) f ( x ) = \sum_{\ell = 1}^{L} \sum_{m = - \ell}^{\ell} a_{\ell, m} \xi_{\ell, m} \, ,
\end{align}
where $\xi_{\ell, m} = \langle f, Y_{\ell, m} \rangle$ is the projection of $f$ onto $Y_{\ell, m}$, is a finite linear combination of mean-zero, independent Gaussian variables. Any such finite sum is itself a Gaussian random variable, so the field $X_a$ is Gaussian even at finite width. If we had chosen the $a_{\ell, m}$ to be independently drawn from a different (non-normal) probability distribution, the sum (\ref{Xa_gaussian_sum}) would not be a Gaussian random variable in general, but Gaussianity at large width could be recovered in certain cases by central limit theorem arguments.

\begin{figure}
  \centering
  \includegraphics[width=\linewidth]{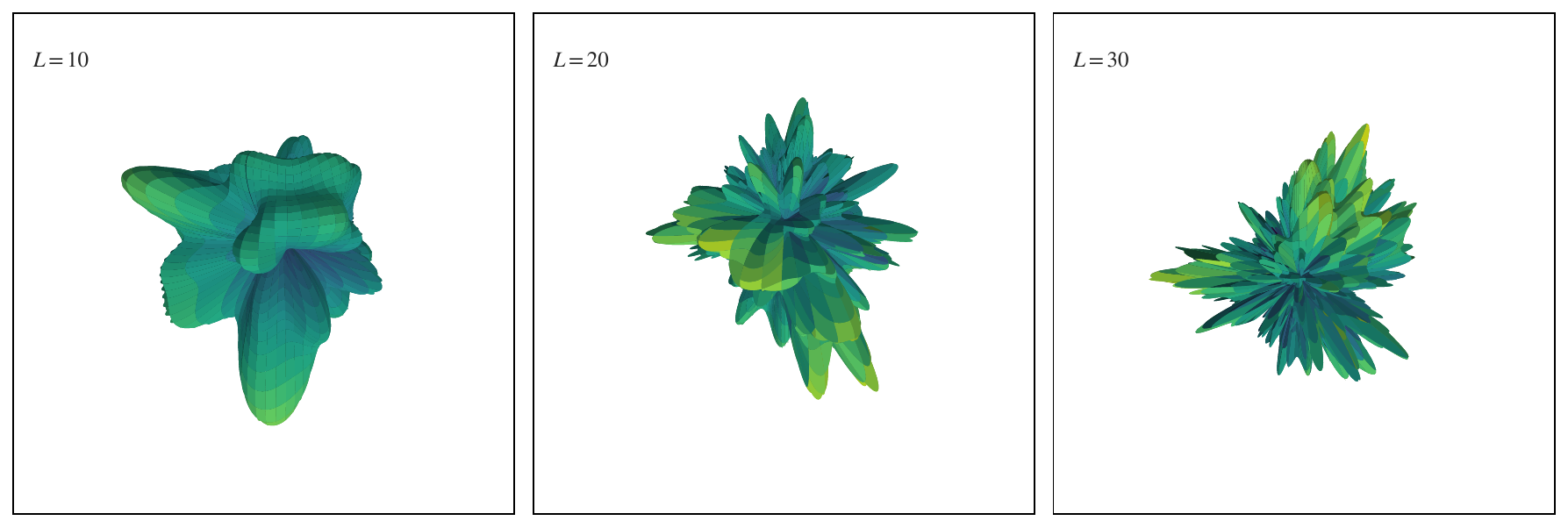}
  \caption{\emph{Distributional Nature of Large Width Liouville Network}. The values of the truncated sum $\sum_{\ell=1}^{L} \sum_{m = - \ell}^{\ell} a_{\ell, m} Y_{\ell, m}$ are shown for increasing values of $L$. The magnitude of the sum is visualized both by the distance from the origin to a point on the surface, and by the color, with yellower tones corresponding to larger distances. As $L$ is increased, more ``delta-function-like'' spikes are seen to develop. This is because the coefficients $a_{\ell, m}$ do not decay sufficiently rapidly for the sum to converge to a smooth function, but the sum does converge almost surely to a distribution on the sphere.}
  \label{fig:schwartz_sphere}
\end{figure}

As we have stressed, in the strict $L \to \infty$ limit where infinitely many spherical harmonics are included, this sum does not converge to a well-behaved function but rather to a distribution. This can be seen visually by examining the behavior of the sum as $L$ is increased, as shown in Figure \ref{fig:schwartz_sphere}. For large $L$, the sum becomes more distributional in character, developing localized spikes much like the Dirac delta distribution.

We incorporate interactions by engineering an appropriate parameter density motivated by the action (\ref{liouville_action}). First note that, since the Ricci scalar of the sphere is constant, one has
\begin{align}
    \int_{S^2} d^2 x \, \sqrt{g} Y_{\ell, m} R = 0 \, ,
\end{align}
for all $\ell \geq 1$, and thus the curvature coupling does not affect the non-zero-mode field $X_a$. However, the zero mode $c$ contributes a term
\begin{align}
    \int_{S^2} d^2 x \, \sqrt{g} \, \frac{Q}{4\pi} c R = 2 Q c
\end{align}
to the action, and thus a factor of $\exp \left( - 2 Q c \right)$ to the weight $e^{-S}$.

The exponential interaction term yields
\begin{align}
    \int_{S^2} d^2 x \, \sqrt{g} \mu e^{2 b c} \, : e^{2 b X} \, : \, = e^{2 b c} M_X \, ,
\end{align}
where
\begin{align}\label{MX_app_defn}
    M_X = \mu \int_{S^2} d^2 x \sqrt{g} \, : e^{2 b X} \, : \, ,
\end{align}
and thus contributes $\exp \left( - e^{2 b c} M_X \right)$ to the weight.

When considering a three-point function of vertex operators,
\begin{align}
    \langle V_{\alpha_1} ( x_1 ) V_{\alpha_2} ( x_2 ) V_{\alpha_3} ( x_3 ) \rangle = \langle e^{2 ( \alpha_1 + \alpha_2 + \alpha_3 ) c} \, : e^{2 \alpha_1 X ( x_1 )} : \, : e^{2 \alpha_2 X ( x_2 )} : \, : e^{2 \alpha_3 X ( x_3 )} : \, \rangle \, ,
\end{align}
we obtain the zero-mode-dependent factor $e^{2 ( \alpha_1 + \alpha_2 + \alpha_3 ) c}$, which we will also account for using the parameter density for $c$. Combining the three contributions to the path integral weight for the zero mode $c$ above, one finds
\begin{align}
    \exp \left( - S_c \right) = \exp \left( - 2 Q c + 2 \sum_i \alpha_i c - e^{2 b c} M_X \right) \, .
\end{align}
It is convenient to change variables to
\begin{align}
    t = e^{2 b c} \, .
\end{align}
In terms of $t$, accounting for all of the contributions in the action, the ``output-space'' description (\ref{three_point_function}) of the Liouville three-point function can then be written in the ``parameter-space'' form
\begin{align}\label{dozz_parameter_space}
    \langle V_{\alpha_1} ( x_1 ) V_{\alpha_2} ( x_2 ) V_{\alpha_3} ( x_3 ) \rangle &= \frac{1}{2b} \Bigg( \int \left( \prod_{\ell \geq 1, m} d a_{\ell, m} P_G ( a_{\ell, m} ) \right) : e^{2 \alpha_1 X_a ( x_1 ) } : \, : e^{2 \alpha_2 X_a ( x_2 ) } : \, : e^{2 \alpha_3 X_a ( x_3 ) } :  \nonumber \\
    &\qquad \qquad \qquad \cdot \int_{0}^{\infty} dt \, \frac{t^{s-1}}{M_X} \hat{P} ( t \mid M_X ) \Bigg) \, ,
\end{align}
where each $P_G ( a_{\ell, m} )$ is a Gaussian distribution,
\begin{align}
    P_G ( a_{\ell, m} ) = \frac{1}{\sqrt{ 2 \pi \sigma^2}} \exp \left( - \frac{a_{\ell, m}^2}{2 \sigma^2} \right) \, , \qquad \sigma^2 = \frac{4 \pi}{\ell ( \ell + 1 ) } \, ,
\end{align}
and
\begin{align}\label{P_normalized}
    \hat{P} ( t \mid M_X ) = M_X \exp \left( -t M_X \right)
\end{align}
is a normalized exponential distribution whose parameter $M_X$ implicitly depends upon the $a_{\ell, m}$. Alternatively, one can write this parameter-space description using the un-normalized distribution $P ( t \mid a_{\ell, m} )$ defined in equation (\ref{P_un_normalized}) by absorbing the additional factors appearing on the second line of (\ref{dozz_parameter_space}). In practice, our NN-FT implementation proceeds by first performing Gaussian draws of the $a_{\ell, m}$, then conditional exponential draws of $t$ following (\ref{P_normalized}), to estimate the parameter-space integral (\ref{dozz_parameter_space}).

This strategy is similar to the one used in \cite{Demirtas:2023fir} to engineer $\phi^4$ theory. One begins with a Gaussian NN-FT, deforms the action by terms which are local in the fields, and then interprets this deformation as a modification of the parameter density which breaks independence. In particular, for the case of LCFT, all draws of $a_{\ell, m}$ determining $X_a$ are independent, and interactions are incorporated through the conditional dependence of the distribution for the zero mode $c$. By construction, this procedure gives rise to a NN-FT with local interactions, as in the $\phi^4$ example.

A few comments are in order. First, to compute the normal-ordering of both the vertex operators $: e^{2 \alpha_i X_a ( x_i )} :$ and $M_X$, we use the formula (\ref{normal_ordered_X}) where the variance $\mathbb{E} \left[ X_{\epsilon} ( x )^2 \right]$ is computed using an ensemble average over numerical draws. In the case of the vertex operators, this is sufficient; for $M_X$, defined by (\ref{MX_app_defn}), we must also integrate over the sphere. In the numerical implementation, we do this by discretizing or ``pixelating'' the sphere at a finite number of grid points, then taking a Riemann sum of normal-ordered quantities on each of the pixels. Second, as is well-known, one can evaluate the contribution from the zero mode integral in closed form as
\begin{align}
    \int_{0}^{\infty} dt \, t^{s-1} e^{- M_X t } = \left( M_X \right)^{-s} \Gamma ( s ) \, ,
\end{align}
but again for completeness we do not use this result in our algorithm. To illustrate a fully numerical NN-FT approach to Liouville theory, we instead sample values of the parameter $t$ from an exponential distribution (in addition to performing Gaussian draws of the $a_{\ell, m}$), and then compute the ensemble average of the integrand of (\ref{dozz_parameter_space}) to experimentally estimate the three-point function.

Another operational detail concerns importance sampling. Due to the form (\ref{normal_ordered_X}) of the normal-ordered vertex operator when $k = 2 \alpha_i$, and the fact that the variance $\mathbb{E} \left[ X_{\epsilon} ( x )^2 \right]$ becomes large as one includes more spherical harmonics, a substantial contribution to the expectation value (\ref{dozz_parameter_space}) arises from parameter draws with small probability but large values of the three-point function. This fact is addressed using the cross-entropy method, which is commonly applied in rare event simulation settings. Specifically, our algorithm begins with a ``warm-up'' phase where field configurations are drawn and the importance score of each draw is computed based on the magnitude of its contribution to the integral defining the three-point function. The algorithm then updates in order to preferentially sample high-scoring configurations and reduce the variance of the estimate to (\ref{dozz_parameter_space}). This component can be viewed as an additional machine learning layer in the algorithm which attempts to identify the optimal sampling strategy to improve the statistics of the simulated three-point function. The importance sampling method repeats for some number of iterations until stabilizing, which in Figure \ref{fig:dozz_prl_plot} was chosen dynamically for each experiment, while in Figure \ref{fig:second_numerics} (to be presented below) was fixed to only five iterations. After this ``warm-up'' phase, the algorithm proceeds to the ``production'' phase where it uses the learned distribution to sample field configurations and estimate (\ref{dozz_parameter_space}).

This procedure does not reproduce the DOZZ formula on the nose, but rather only up to an overall multiplicative factor. One way to think about this is the following. The formula (\ref{DOZZ_C}) includes a prefactor depending on parameters in the Lagrangian. For instance, let us focus on the factor $\mu^{-s}$. Because our algorithm implicitly imposes a UV cutoff $\Lambda$ in multiple steps -- both by explicitly pixelating the sphere in the computation of $M_X$, and by truncating to finitely many modes for $X_a$ -- the numerical correlation function that we are computing should really depend upon a renormalized coupling $\mu_{\text{eff}} ( \Lambda )^{-s}$ rather than the bare quantity $\mu^{-s}$. Instead of attempting to predict this renormalized coupling directly, we have chosen to perform a fit of the overall $\alpha$-independent prefactor in $C ( \alpha_1, \alpha_2, \alpha_3 )$ in each run of our algorithm for fixed parameters $\mu, b$. Although this overall constant is obtained from a fit, we emphasize that the remaining trend across different values of the $\alpha_i$ agrees well with the DOZZ formula.

We will not undertake a detailed sensitivity analysis in this work, but let us make a few basic comments about sources of error in this algorithm. First, one chooses a finite number $L$ of spherical harmonics to include in the sum defining the architecture $X_a$, and the accuracy naturally appears to increase for larger $L$. We choose a finite number $N$ of sample field configurations, introducing Monte Carlo error for quantities which is expected to scale as $\frac{1}{\sqrt{N}}$. In order to numerically estimate the value of the normal-ordered quantity $M_X$, we pixelate the sphere at some fixed number of points, and the error of the algorithm again decreases as this pixelation is made finer. As we have mentioned, one step of our implementation involves using importance sampling to better estimate the contributions from the normal-ordered expressions $: e^{2 \alpha_i X ( x_i )} :$, and the error decreases as more iterations of this importance sampling step are performed. Finally, there are two physical conditions on the $\alpha_i$. The first is the Seiberg bound,
\begin{align}\label{seiberg_bound}
    \alpha_i \leq \frac{Q}{2}
\end{align}
for all $i$, while the second is
\begin{align}
    \sum_i \alpha_i > Q \, ,
\end{align}
which is needed for the Liouville path integral to converge. As one approaches either of these bounds, the error of our algorithm increases because the three-point function being computed becomes pathological. In particular, the DOZZ prediction for the three-point function is typically zero when the Seiberg bound is saturated. Some data points which are close to the Seiberg bound and thus have larger error are presented in Figure \ref{fig:second_numerics}.

\begin{figure}
  \centering
  \includegraphics[width=\linewidth]{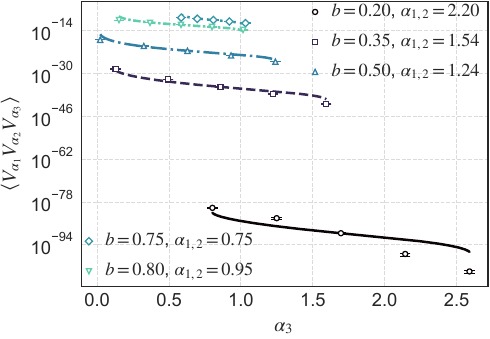}
  \caption{We compare the results of our NN-FT simulation to the DOZZ formula for values of the three-point function which are smaller than those shown in Figure \ref{fig:dozz_prl_plot}. Again, in each experiment, two values of the $\alpha_i$ are held fixed while the third is varied. Some of these data points are close to the Seiberg bound (\ref{seiberg_bound}), where the DOZZ formula generically has a zero and the techniques used to define Liouville correlators become more subtle. We draw $N = 75,000$ sample field configurations, pixelate the sphere at $128$ points of longitude and $64$ points of latitude, and perform $5$ iterations of importance sampling. For these values of parameters, the errors are larger and several of the DOZZ formula predictions do not lie within the error bars representing the standard deviation across experiments.}
  \label{fig:second_numerics}
\end{figure}

\end{document}